\documentclass[10pt,twocolumn,
amssymb, nobibnotes, notitlepage,nofootinbib aps, prd]{revtex4}

\usepackage{amsmath}
\usepackage{amssymb}
\usepackage{amsthm}
\usepackage[pdftex]{color}
\usepackage[pdftex,colorlinks,citecolor=blue,linkcolor=blue,urlcolor=blue]{hyperref} 
\usepackage{graphicx}
\usepackage{dcolumn} 
\usepackage{bm} 

\usepackage{longtable}

\usepackage{ulem}   
\usepackage{comment} 
\usepackage{datetime}

\usepackage{tabularx}
\usepackage{float}
\restylefloat{table}

\newcommand{\beq}{\begin{equation}}
\newcommand{\eeq}{\end{equation}}
\newcommand{\bea}{\begin{eqnarray}}
\newcommand{\eea}{\end{eqnarray}}
\newcommand{\barr}{\begin{array}}
\newcommand{\earr}{\end{array}}

\long\def\begincomment#1\endcomment{}

\begin{document}
	
	
	
\title{Constructing the low-temperature phase diagram for the ‘$2+p$’-quantum spin glass using the nonperturbative renormalization group }

\author{Ixandra Achitouv}
\email[]{ixandra.achitouv@cnrs.fr}
\affiliation{Institut des Syst\`emes Complexes ISC-PIF , CNRS, 113 rue Nationale,
Paris, 75013, France.}

\author{Vincent Lahoche}
\email[]{vincent.lahoche@cea.fr}
\affiliation{Université Paris Saclay, \textsc{Cea}, Gif-sur-Yvette, F-91191, France}

\author{Dine Ousmane Samary}
\email[]{dine.ousmanesamary@cea.fr}
\affiliation{Université Paris Saclay, \textsc{Cea}, Gif-sur-Yvette, F-91191, France}
\affiliation{Faculté des Sciences et Techniques (ICMPA-UNESCO Chair)\protect\\
Université d'Abomey-Calavi, 072 BP 50, Bénin}

\author{Parham Radpay}
\email[]{parham.radpay@cea.fr}
\affiliation{Université Paris Saclay, \textsc{Cea}, Gif-sur-Yvette, F-91191, France}

\begin{abstract}
In this paper, we use a nonperturbative renormalization group approach to construct the dynamical phase space of a quantum spin glass in the large $N$ limit. The disordered Hamiltonian is of ``$2 + p$" type, and we perform a coarse-graining procedure over the Wigner spectrum for the matrix-like disorder. The phase space reconstruction relies on phase transitions derived from the Luttinger-Ward functional, which accounts for interactions that are forbidden by perturbation theory. Various phases are identified, characterized by large correlations between replicas and/or the breaking of time translation symmetry.
\end{abstract}
	
\maketitle

\section{Introduction}

Since its inception, research on spin glasses has provided valuable insights into disordered materials and the glassy state, while also generating new methods for tackling a wide range of problems, from computer architecture to quantum gravity to economics \cite{Ellman,SG1,SG2,SG3,Mezard3,SG4,SG5}. Glassy systems, in general, are characterized by their slow dynamics and non-equilibrium effects, especially at low temperatures \cite{Dotsenko,Castellani,Mezard,Dominicis,Agliari,Leuzzi}. Typically, classical statistical mechanics is employed to describe spin glasses, as the typical energy scale, $k_B T_c$, near the transition temperature is usually large enough that quantum effects play a negligible role. However, in some cases, the critical temperature $T_c$ depends on external parameters and can be reduced to arbitrarily low values, allowing quantum effects, such as tunneling, to suppress the glass transition. For concrete examples from experimental physics, the reader may refer to \cite{Ellman, Rosenbaum2, Vollmer} and the references therein.

\medskip

Quantum effects on infinite-range models of spin glasses have been studied for several decades \cite{QSG3,QSG4,QSG5}. Many models assume that quantum fluctuations (particularly tunneling) play a role analogous to classical thermal fluctuations. A typical model incorporates a transverse magnetic field, which does not commute with the Ising interaction \cite{QSG6}. Experimentally, quantum fluctuation effects have been observed in $\text{Li}\text{Y}_{1-x} \text{Ho}_x \text{F}_4$ \cite{Ellman}. In other cases, the role of quantum fluctuations is significantly different, particularly in cuprate physics and generally when there is no trivial quantum ground state. These models, which often lack an analogy to the Ising model, generalize the Sherrington-Kirkpatrick model to a quantum Heisenberg model with special unitary spin rotation symmetry. A paradigmatic example of this approach is the Sachdev-Ye-Kitaev (SYK) model, which is characterized by extensive zero-temperature entropy and has found applications in quantum gravity \cite{SG3,Maldacena}. Another approach to understanding quantum aspects involves models that describe a quantum particle moving through a random potential, akin to a quantized version of the classical spherical $p$-spin model \cite{Cugliandolo1,Cugliandolo2,Cugliandolo3}. This class of models is the focus of the present paper.

\medskip

Recent applications of spin glass physics, and the understanding of the underlying quantum effects, have increasingly focused on classical optimization problems by exploiting the tunneling effect \cite{QSG1,QSG2,QSG6}. Among these challenging classical problems are those involving "structured" disorders in spin glasses \cite{Mezard2}, such as the analysis of matrix and tensor principal components \cite{PCA1,PCA2}. This article follows the recent line of work \cite{Lahoche1,Lahoche2,Lahoche3,Lahoche6}, which uses the renormalization group to explore an analogous model corresponding to a quantum particle evolving in a rough potential, realized by one or more random tensors. One of the goals of our investigation is to develop reliable RG techniques for analyzing structured quantum signals.

\medskip

In this paper, continuing from \cite{Lahoche1,Lahoche2,Lahoche3,Lahoche6}, we aim to construct the full phase diagram in the symmetric phase for this model as a benchmark for the functional renormalization group approach. In particular, we focus on phase transitions induced by metastable states corresponding to two-point interactions that are forbidden by large $N$ perturbation theory, leading to different phases characterized by macroscopic correlations between replicas and time-translation symmetry breaking. Our method employs the 2PI formalism, focusing on the large $N$ expansion of the Luttinger-Ward functional \cite{Blaizot,Dominicis}, as we discussed in \cite{Lahoche6}.

\section{Theoretical background}

\begin{figure}
\begin{center}
\includegraphics[scale=0.9]{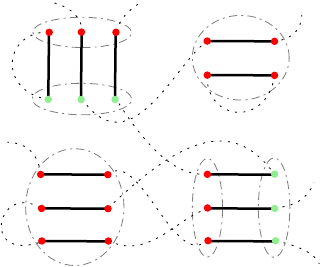}
\end{center}
\caption{A typical Feynman (hyper-) graph contributing to the $6$-point function. }\label{Feynman}
\end{figure}

The model we consider is analogous to the quantum mechanical problem of a single particle in $\mathbb{R}^N$ moving through a random energy landscape. The Hamiltonian matrix elements in the generalized position basis are given by:
\begin{equation}
\hat{\mathcal{H}}(\textbf{x},t):=-\frac{\hbar^2}{2 m_0} \frac{\partial^2}{\partial \textbf{x}^2} + U(\textbf{x}^2) + V_{J,K}(\textbf{x})\,,
\end{equation}
where $m_0$ is the physical mass of the particle, $U$ is some polynomial function with argument the single $O(N)$ invariant $\textbf{x}^2:=\sum_{i=1}^N x_i^2$, and:
\begin{equation}
V_{J,K}(\textbf{x}):=\frac{1}{2} \sum_{i,j} K_{ij} x_i x_j+\sum_{i_1\leq \cdots \leq i_p} J_{i_1\cdots i_p}x_{i_1}\cdots x_{i_p}\,.
\end{equation}
In this equation $K$ and $J$ are quenched random couplings, $K$ is a Wigner matrix \cite{RMT} of size $N$ and variance $\sigma^2$, and $J$ is a Gaussian random tensor with zero mean and variance:
\begin{equation}
\overline{J_{i_1\cdots i_p}J_{i_1^\prime \cdots i_p^\prime}}=\left(\frac{\kappa^2 p!}{N^{p-1}}\right) \,\prod_{\ell=1}^p \delta_{i_\ell i_\ell^\prime}\,.\label{averageJ}
\end{equation}
We use the notation $\overline{X}$ for the average over rank $p$ disorder distribution. In the large $N$ limit, the spectrum for the matrix $K$ converge weakly toward the \textit{Wigner semi-circle} distribution:
\begin{equation}
\mu_W(\lambda):=\frac{\sqrt{4\sigma^2-\lambda^2}}{2\pi \sigma^2}\,.\label{Wigner}
\end{equation}
We consider the quantum particle issue in contact with thermal bath with temperature $\beta^{-1}$, whose partition function,
\begin{align}
\mathcal{Z}_\beta:= \int d\textbf{x}\, \langle \textbf{x} \vert e^{-\beta \hat{\mathcal{H}}} \vert \textbf{x} \rangle\,,
\end{align}
which can be rewritten using Feynman path integral \cite{Feynman}:
\begin{equation}
\mathcal{Z}_{\beta}[\textbf{L}]\equiv \int [\mathcal{D} x(t)]\, e^{-\frac{1}{\hbar} S_{\text{cl}}[\textbf{x}(t)]+\frac{1}{\hbar} \int dt \sum_{k=1}^N L_k(t) x_k(t)}\,,\label{pathintegralZ}
\end{equation}
such that $\mathcal{Z}_{\beta}[\textbf{L}=0]\equiv \mathcal{Z}_{\beta}$ and where the classical action is:
\begin{equation}
S_{\text{cl}}[\textbf{x}(t)]:=\int_{-\beta/2}^{\beta/2} dt \left(\frac{1}{2}\dot{\textbf{x}}^2+V_{J,K}({\textbf{x}})+U({\textbf{x}}^2)\right)\,,\label{classicaction}
\end{equation}
provided with periodic boundary conditions $\textbf{x}(t)=\textbf{x}(t+\beta)$, implying that Fourier frequencies are quantified $\omega_n=2\pi n/\beta$. In the Gauge where the matrix like disorder is diagonal and assuming the probability distributions for $K$ and $J$ are $O(N)$ invariants, the kinetic kernel is, in the Fourier space:
\begin{equation}
\mathcal{K}=\omega^2+\lambda_\mu+2 U'(0)=: \omega^2+p_\mu^2+m^2\,,
\end{equation}
where the \textit{generalized momentum} $p_\mu^2:=\lambda_\mu+2\sigma$ is a positive quantity in the limit $N\to \infty$ (meaning that the probability of obtaining a negative value approaches zero), and $m^2:=2 U'(0)-2\sigma$.
We denote $\rho(p^2)$ as the large-$N$ distribution of the generalized momentum, derived from $\mu_W$. The fact that the random spectrum of the matrix $K$ converges to a deterministic law suggests that $K$ and $J$ should be treated differently.
By fixing $J$, all perturbative amplitudes indexed by Feynman diagrams can be computed using the effective deterministic distribution $\mu_W$, rather than summing explicitly over the discrete random spectrum of $K$. The corresponding Feynman graphs define an alternative field theory that depends only on the distribution $\mu_W$, schematically:
\begin{equation}
\mathcal{Z}_{\beta}[\textbf{L}] \to \tilde{\mathcal{Z}}_{\beta}[\mu_W,\textbf{L}]\,,
\end{equation}
where $\tilde{\mathcal{Z}}\beta$ does not depend on $K$ but still depends on $J$. The construction of the averaging over $J$ becomes more complicated when we consider it in the quenched regime. The most popular method uses the replica trick, where the averaging is done over $n$ copies of $\tilde{\mathcal{Z}}$ before sending $n \to 0$. This analytical continuation reveals the phenomenon of replica symmetry breaking (RSB) \cite{Mezard,Castellani}. In the renormalization literature, a different approach is considered \cite{Tarjus1,Tarjus2,Tarjus3,Tarjus4,Dupuis}, which focuses on the moments of the random free energy $\ln \tilde{\mathcal{Z}}{\beta}[\mu_W,\textbf{L}]$, which can be obtained from the averaging of the replicated partition function:
\begin{equation}
\boxed{\bar{Z}_\beta[\mu_W,\{\textbf{L}_\alpha\}]:= \overline{\prod_{\alpha=1}^n \tilde{\mathcal{Z}}_{\beta}[\mu_W,J,\textbf{L}_\alpha]}\,,}\label{averagedZ}
\end{equation}
where in this equation, $n$ is some arbitrary integer, and replica symmetry is \textit{explicitly broken} because the source fields $\textbf{L}_\alpha$ are different. The classical action for the averaged replicated theory is:
\begin{align}
\nonumber\overline{S_{\text{cl}}}&[\{\textbf{x}_\alpha\}]:=\sum_{\alpha}S_{\text{cl}}[\textbf{x}_\alpha(t),J=0,K]\\
&\quad -\frac{\kappa^2 N}{2\hbar}\int_{-\beta/2}^{+\beta/2} d t \, d t^\prime\sum_{\alpha,\beta}\,  \left(\frac{\textbf{x}_\alpha(t)\cdot \textbf{x}_\beta(t^\prime)}{N}\right)^p\,.\label{classicalaveraged}
\end{align}
Note that we will set $\hbar = 1$ and $p = 3$ up to this point. The Feynman amplitudes in perturbation theory are labeled by hypergraphs rather than ordinary Feynman graphs \cite{Hyper}, as Figure \ref{Feynman} illustrates. The graphical rules are as follows: Dots materialize fields, and all the fields enclosed by a dash-dotted bubble interact simultaneously. Moreover, in different local components, fields do not share the same replica and are materialized by different colors. For instance:
\begin{equation}
\vcenter{\hbox{\includegraphics[scale=0.8]{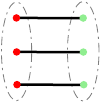}}}\,=\, \int dt dt^\prime \sum_{\alpha,\beta=1}^n\, (\textbf{x}_\alpha(t) \cdot \textbf{x}_\beta(t^\prime))^3\,.
\end{equation}
Moreover, Wick contraction with the bare propagator are materialized with dotted edges.

We aim to construct an RG flow by performing a coarse-graining over the Wigner spectra, in accordance with the method we proposed in \cite{Lahoche1,Lahoche2,Lahoche3,Lahoche5}, which we will summarize here. We focus on the functional formalism developed by Wetterich \cite{Wetterich1,Wetterich2,Wetterich3,Delamotte}, and modify the averaged classical action \eqref{classicalaveraged} by adding a regulator $\Delta S_k$ that suppresses IR modes with $p^2 < k^2$ from long-range physics:
\begin{equation}
\Delta S_k[\{\textbf{x}_\alpha\}]:= \frac{1}{2}\sum_{\alpha=1}^n \sum_{\mu=1}^N \int dt\, x_{\alpha \mu}(t) R_k(p_\mu^2)x_{\alpha i}(t)\,.
\end{equation}
In this paper we will focus on the slightly modified Litim regulator, $R_k(p^2):=f(k)(k^2-p^2)\theta(k^2-p^2)$, where the factor $f(k):=4\sigma/(4\sigma-k^2)$ accommodates with the compact nature of the Wigner spectrum. 

In \cite{Lahoche1,Lahoche2,Lahoche3,Lahoche6}, we showed that the non-local sextic coupling arising from averaging over the rank $p$ disorder does not renormalize when $p > 2$. Furthermore, RG flow equations derived from the standard one-particle irreducible (1PI) formalism exhibit finite-scale singularities for sufficiently large $\kappa$. As we investigated in our previous works, these singularities arise because metastable states, which dominate the RG flow for strong disorder, correspond to interactions forbidden by perturbation theory and dominate the flow from some finite scale \cite{Delamotte,Dupuis,Tarjus1}. Taking these interactions into account cancels the singularities.

In this paper, we aim to use these metastable states to construct the phase space of the system, with different regions corresponding to the various phase transitions occurring along the RG flow. We expect that these phase transitions correspond to metastable states that break time-translation symmetry and/or couple replicas. In both cases, we assume the order parameter is the two-point function, in accordance with the quantum glassy system literature \cite{Dominicis,Baardewijk}.
\medskip

In the case where the order parameter is the $2$-point function, the two-particle irreducible (2PI) formalism applies. The explicit construction requires introducing sources for the $2$-point functions. Within the 2PI formalism, the fundamental quantity is the replicated 2PI effective action $\Gamma_k[{\textbf{M}\alpha },{G\alpha}]$, which depends on the 1-point functions $\textbf{M}\alpha$ and on the 2-point functions $G\alpha$, and is defined as \cite{Blaizot,Gurau}:
\begin{align}
\Gamma_k[\{\textbf{M}_\alpha \},G]=\frac{1}{2} \mathrm{Tr}\ln G^{-1}+\frac{1}{2} \mathrm{Tr} \, G_0^{-1} G + \Phi[G]\label{2PIfunctional}
\end{align}
where $\mathrm{Tr}$  means sum over momenta, frequencies and replica, and the bare propagator $G_0$ (diagonal in the replica space) is defined as:
\begin{equation}
G_0(\omega^2,p^2):=\omega^2+p^2+m^2+R_k(p^2)\,.\end{equation}
The last piece in the definition \eqref{2PIfunctional}, $\Phi[G]$ is the so-called Luttinger-Ward functional and expands in term of 2PI diagrams, which in particular determines the gap equation:
\begin{equation}
\Sigma=-2\, \frac{\delta \Phi[G]}{\delta G}\,,\label{gapeq}
\end{equation}
where $\Sigma$ is the standard self energy. The solution of this equation is noting but the so-called Dyson equation, and we denote it as $G_k$.
In the large $N$ limit, the functional $\Phi$ can be computed exactly \cite{Lahoche5,Gurau}, graphically, and for a sextic theory:
\begin{equation}
\Phi[G]\,=\,\vcenter{\hbox{\includegraphics[scale=0.8]{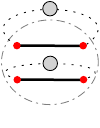}}}\,+\, \vcenter{\hbox{\includegraphics[scale=0.8]{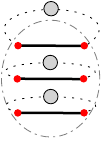}}}\,+\, \vcenter{\hbox{\includegraphics[scale=0.8]{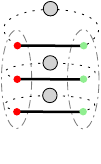}}}\,,
\end{equation}
where the dotted edges with gray discs represent the propagator $G$, the notation indicating that graphs are computed as Feynman amplitudes, replacing the bare propagator with $G$. The 2PI formalism can easily be connected with the standard 1PI \cite{Blaizot}. Indeed, denoting $f_k := \Gamma_k[{\textbf{M}\alpha }, G_k]$ as the effective action \textit{on shell}, i.e., for $G = G_k$, and since $\Gamma_k[{\textbf{M}\alpha }]$ depends on $k$ only through $G_0$, we obtain:
\begin{equation}
\dot{f}_k=\frac{1}{2}\mathrm{Tr}\, \dot{R}_k \, G_k\,,
\end{equation}
which is formally equivalent to the standard 1PI flow equation, known as the Wetterich equation. Approximate solutions to this equation can be constructed in the symmetric phase using vertex expansion, which is more suitable and tractable for this kind of non-local field theory than other standard tools \cite{Lahoche1,Lahoche4}. The standard approximation scheme in such cases consists of projecting both sides onto a restricted region of the full theory space, spanned by some ansatz for $\Gamma_k$ \cite{Wetterich1,Wetterich2,Wetterich3}. For our purposes, the ansatz we consider is a multi-local expansion \cite{Lahoche4,Dupuis,Tarjus1}. Assuming we consider only local and bilocal contributions:
\begin{align}
\nonumber f_k&=\frac{1}{2}\int dt \sum_{\mu,\alpha} M_{\mu,\alpha}(t)(-\partial_t^2+p_\mu^2+u_2)M_{\mu,\alpha}(t)\\\nonumber
&+\sum_{n=2}^\infty\int dt \sum_{\mu,\alpha} \frac{(2\pi)^{n-1}u_{2n}}{(2n)!N^{n-1}}\, \bigg(\sum_\mu M_{\mu,\alpha}^2(t) \bigg)^n\\
&+\frac{(2\pi)^2\tilde{u}_6}{6!N^2}\int dt dt^\prime \sum_{\alpha,\beta}\, \bigg(\sum_\mu M_{\mu,\alpha}(t) M_{\mu,\beta}(t^\prime) \bigg)^3\,,\label{truncationGamma}
\end{align}
which, in turn, corresponds to the on-shell approximation:
\begin{equation}
G_k(\omega,p^2):=\frac{1}{\omega^2+p^2+u_2+R_k(p^2)}\,.
\end{equation}
Moreover, $\tilde{u}_6\equiv -6! \kappa^2/(8 \pi^2)$ does not renormalizes. Flow equations for couplings can be derived explicitly from the vertex expansion \eqref{truncationGamma}. However, and in contract with ordinary field theory, it does not exist a suitable rescaling of couplings making the flow equations autonomous; assuming $u_2\ll k^2$, and for $k$ small enough, one get for instance \footnote{There in an overall factor $2\pi$ regarding the beta-functions considered in \cite{Lahoche1}, due to the convention used in the reference, which multiplied the flow equations by such a factor implicitly because of the definition of canonical dimension. This does not, however, qualitatively change the results.}:
\begin{equation}
\dot{\bar{u}}_4\approx-\mathrm{dim}_{4}\bar{u}_4- \frac{\bar{\tilde{u}}_6 }{15\pi}-\frac{\bar{u}_6 }{30}+ \frac{\bar{u}_4^2}{6}\,,\label{flowu4}
\end{equation}
\begin{equation}
\dot{\bar{{u}}}_6\approx -\mathrm{dim}_{6}\, {\bar{{u}}}_6+\frac{144\bar{u}_4\bar{u}_6}{5}+\frac{8\bar{u}_4\bar{\tilde{u}}_6}{5\pi} -\frac{5\bar{u}_4^3}{9} \,,\label{flowu6}
\end{equation}
where:
\begin{equation}
\boxed{\mathrm{\dim}_{2n}(k):= (n-1)\frac{k^2}{k^2-4}+2(2-n) \,,}
\end{equation}
and:
\begin{equation}
\bar{u}_{2n}=u_{2n}\, \frac{1}{k^2}\, \left(\int dp^2\, \frac{\rho(p^2) \dot{R}_k(p^2)}{(p^2+R_k(p^2))^{3/2}}\right)^{n-1}\,.\label{rescalingGaussian}
\end{equation}

Remark that to derive the flow equations before, we neglected some numerical factors whose are numerically of order $1$ for $k$ small enough -- see \cite{Lahoche1,Lahoche6}.
\medskip

We consider a Ginzburg-Landau approach \cite{Dupuis2}, assuming that near the transition, metastable states correspond to small enough effective interactions. Note that this condition makes sense for a continuous, i.e., second-order phase transition, but not necessarily for discontinuous phase transitions, as we found in our previous works \cite{Lahoche1,Lahoche2,Lahoche3,Lahoche6}. This point is a weakness of our approach, which we intend to improve in the continuation of our work. The principle is as follows. We assume that the self-energy $\Sigma$ splits into two contributions:
\begin{equation}
\Sigma=\Sigma_N+\gamma
\end{equation}
where here, $\Sigma$, $\Sigma_N$ and $\gamma$ are $n\times n$ matrices. The self energy $\Sigma_N$ is what we expect from the large $N$ perturbation theory, it is diagonal in the replica space, and almost independent from frequency: $(\Sigma_N)_{\alpha\beta,\mu\nu}(\omega,\omega^\prime)\approx -u_2(k) \delta_{\alpha\beta} \delta(\omega+\omega^\prime)$. The contribution $\gamma$, forbidden by perturbation theory is expected to be small enough. Hence, on shell expanding the state (gap) equation \eqref{gapeq} leads formally to:

\begin{align}
\nonumber{\gamma}&\,=\, \vcenter{\hbox{\includegraphics[scale=0.6]{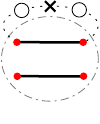}}}\,+\,\vcenter{\hbox{\includegraphics[scale=0.6]{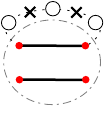}}}\,+\,2\times \vcenter{\hbox{\includegraphics[scale=0.6]{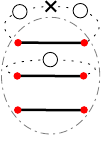}}}\\\nonumber
&+\, \vcenter{\hbox{\includegraphics[scale=0.6]{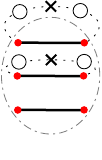}}}\,+\,2\times  \vcenter{\hbox{\includegraphics[scale=0.6]{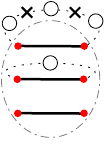}}}\,+\, 2\times  \vcenter{\hbox{\includegraphics[scale=0.6]{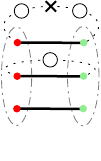}}}\\
&+\, 2\times  \vcenter{\hbox{\includegraphics[scale=0.6]{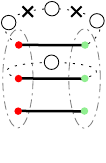}}}\,+\,\vcenter{\hbox{\includegraphics[scale=0.6]{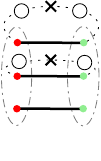}}}\,+\,\mathcal{O}(\gamma^3)\,,
\end{align}
where white bubbles represent the effective propagator in the normal phase, and here crosses are $\gamma$ insertions. Note that some diagrams involved in the sum must be discarded, depending on the explicit nature of the operator we consider. From our previous works cited above, we will consider four different cases (dynamical vs. equilibrium ergodicity breaking — see also \cite{QSG6,Cugliandolo1,Cugliandolo2}):
\begin{align}
(\gamma_1)_{\alpha\beta,\mu \nu}(\omega,\omega^\prime)&=\, -\Delta_1 \delta_{\mu \nu}\delta_{\alpha\beta}^\bot \delta(\omega+\omega^\prime)\\
(\gamma_2)_{\alpha\beta,\mu \nu}(\omega,\omega^\prime)&=\, -\Delta_2\delta_{\mu\nu}\delta_{\alpha\beta} \\
(\gamma_3)_{\alpha\beta,\mu \nu}(\omega,\omega^\prime)&=\, -\Delta_3 \delta_{\mu\nu}\delta_{\alpha\beta}^\bot(\delta(\omega)+\delta(\omega^\prime))\\
(\gamma_4)_{\alpha\beta,\mu \nu}(\omega,\omega^\prime)&=\,-\Delta_4 \delta_{\mu\nu}\delta_{\alpha\beta}^\bot \,,
\end{align}
where $\delta_{\alpha\beta}^\bot:=1-\delta_{\alpha\beta}$ avoids back reaction on the leading order 1PI flow for the mass $u_2$. These different interactions make two thing: they couple replica and break time translation symmetry. The state equation then split into 4 equations we identify from they specific nature:
\begin{equation}
\Delta_i=F_{i}(\Delta_i)\,.
\end{equation}
Before we compute it exactly, $F_{i}(\Delta)$ looks as a gradient flow $\Delta_i-F_{i}(\Delta_i):=\partial_{\Delta_i} U(\Delta_i)$, for some effective potential $U_i(\Delta_i)$. The explicit construction of the potential follows the method we considered in \cite{Lahoche2,Lahoche6} and require to go until order $\Delta^4$. Explicitly, we get:
\begin{align}
 U_1(\Delta_1)&=\frac{1}{2}\Delta^2_1-\frac{1}{3}a_1 \Delta^3_1-\frac{1}{4} a_2\Delta^4_1\,,\label{potential1}\\
U_2(\Delta_2)&=\frac{1}{2}(1-b_1)\Delta^2_2-\frac{1}{3}b_2 \Delta^3_2\,,\label{potential2}
\\
U_3(\Delta_3)&=\frac{1}{2}\Delta^2_3+\frac{1}{3}c_1 \Delta^3_3+\frac{1}{4} c_2\Delta^4_3\,,\label{potential3}\\
U_4(\Delta_4)&=\frac{1}{2}\Delta^2_4+\frac{1}{3}d_1 \Delta^3_4+\frac{1}{4} d_2\Delta^4_4\,,\label{potential4}
\end{align}
where, explicitly:

\begin{align}
a_1:=&-\frac{\tilde{u}_6}{15}\, \int \rho(p^2) dp^2 \rho(q^2) dq^2\, J_{2,2,0}\,,\\
a_2:=&\frac{2(n-1)\tilde{u}_6}{15}\, \int \rho(p^2) dp^2 \rho(q^2) dq^2\, J_{3,2,0}\,,
\end{align}

\begin{align}
\nonumber b_1:=&-\left(\frac{u_4}{3}+\frac{2u_6}{15} L_1 \right)\,L_2\\
&-\frac{2\tilde{u}_6}{15} \int \rho(p^2)dp^2 \rho(q^2) dq^2 J_{2,1,0}\\\nonumber 
b_2&:=\left(\frac{u_4}{3} + \frac{2 u_6}{15} L_1\right)\mathrm{Tr}\,I_{2,0}I_{1,0}+\frac{\tilde{u}_6}{15} \mathrm{Tr}\, J^2_{1,1,0}\\\nonumber
&+\frac{2\tilde{u}_6}{15}\int \rho(p^2) dp^2 \rho(q^2)dq^2 I_{1,0}J_{2,1,0}\\
&+\frac{u_6}{15}\int \rho(p^2) dp^2 \rho(q^2)dq^2 K(p^2,q^2)\,\,,
\end{align}
and 
\begin{align}
c_1:=&\frac{\tilde{u}_6}{15}\, \Big(\int \rho(p^2) dp^2 G_k(0,p^2)\Big)^2\\\nonumber 
c_2:=&-\frac{2\tilde{u}_6}{15}\, \int \rho(p^2) dp^2 \rho(q^2) dq^2\,G_k^3(0,p^2) G_k^2(0,q^2)\,,
\end{align}

\begin{align}
d_1:=& \frac{\tilde{u}_6}{15} \mathrm{Tr}\, J^2_{1,1,0}\\
d_2:=&-\frac{2 (n-1)\tilde{u}_6}{15} \mathrm{Tr} \,(J_{1,1,0}^2 I_{1,0})\,.
\end{align}
where for $a_2$ we took into account the order $1$ effect on the 1PI flow for $u_2$ \cite{Lahoche3}, and where we used the definitions:

\begin{align}
&\nonumber I_{m,n}:= \int d\omega\, G_k^m(\omega) \omega^{2n}\\
&=C(n,m) (p^2+R_k(p^2)+u_2)^{-n+m-\frac{1}{2}}\,,
\end{align}
where the numerical factor depending on $n,m$ is $C(n,m):=\Gamma \left(n+\frac{1}{2}\right)  \Gamma \left(m-n-\frac{1}{2}\right)/\Gamma(m)$, $L_n:=\int \rho(p^2) d p^2 I_{n,0}$, 
\begin{equation}
J_{m,n,p}:=\int d\omega\, G_k^m(\omega) G_k^n(\omega) \omega^{2p}\,,
\end{equation}
and:
\begin{align}
\nonumber K(p^2,q^2):=\int & d \omega d\omega' d\omega" G_k(\omega,p^2) G_k(\omega',p^2)\\
& \times G_k(\omega",q^2)  G_k(\omega+\omega'+\omega",q^2)\,.
\end{align}
We furthermore used the following conventions: $J_{m,n,p}$ depends on two generalized momenta, the one of the $m$ propagators and the one of the $n$ propagators. Everywhere, $\mathrm{Tr}$ is over generalized momenta; explicitly we have for instance $\mathrm{Tr}\, J^2_{1,1,0}\equiv \int \rho(p^2) \rho(q^2) dp^2 dq^2 J_{1,1,0}(p^2,q^2) J_{1,1,0}(q^2,p^2)$, where we indicated explicitly the dependency over generalized momenta. In the same way, $I_{n,m}$ depends on a single generalized momenta, and the notation $I_{n,m}J_{k,l,r}\equiv I_{n,m}(p^2)J_{k,l,r}(p^2,q^2)$.

\section{Results and conclusion}
\begin{figure} \begin{center} \includegraphics[scale=0.4]{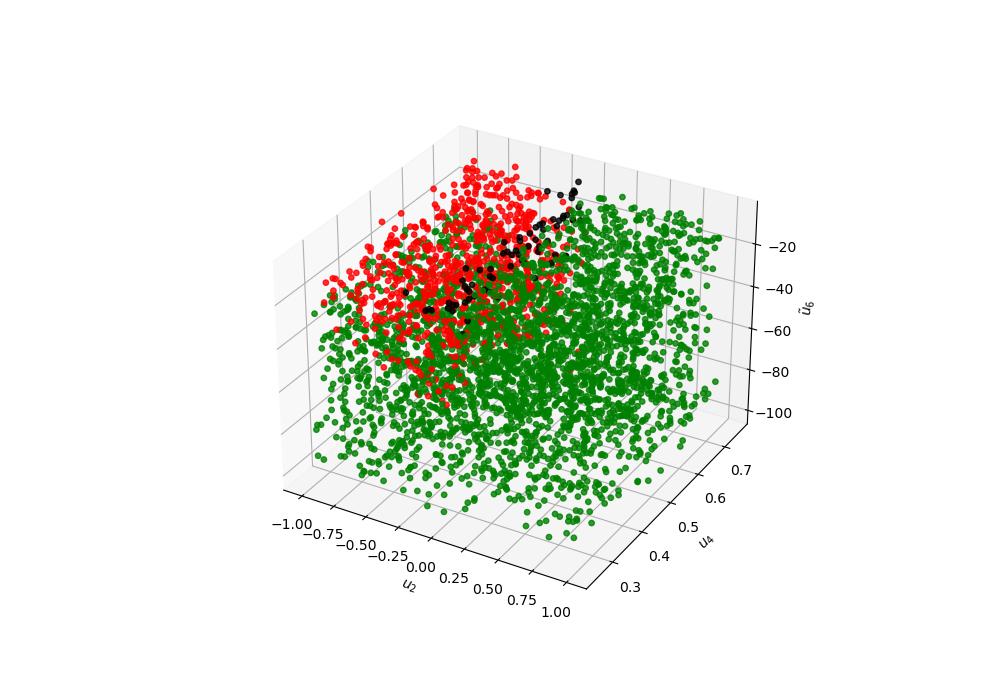} \end{center} \caption{Numerical phase space in the vicinity of the Gaussian region for local couplings: Red dots represent singular trajectories where $U_2$ is broken. Green dots represent singular trajectories where $U_3$ is broken. Black dots represent convergent trajectories.}\label{figPlot} \end{figure}

 The study of the behavior of the potentials considered previously allows us to characterize the type of transition hidden behind the finite-time singularity of the RG flow and to consider a reconstruction of the effective phase space of the model. For this, we will consider a certain UV scale, $k_0=1.999$, and we will focus on a region of the phase space close to the Gaussian point in the local couplings. In \cite{Lahoche6}, we showed an almost clear separation in the full phase space between a singular regime and a regime where the flow converges, and we will focus here only on the singular region. Our idea is as follows: we put the four potentials in competition, for a given initial condition, and the first of the four that exhibits a phase transition will assign a specific color to the initial condition in question. The numerical code systematizing this process can be found via the link \url{https://github.com/ParhamRadpay/2-p-Spin-Glass-Phase-Space}, and the results are summarized in Figure \ref{figPlot}. Each dot corresponds to some initial condition, and the color means the following:

\begin{enumerate} \item Black dots represent convergent trajectories (without finite-scale singularity). \item Red dots represent trajectories having a finite-scale singularity such that potential $U_2$ exhibits a second-order phase transition along it. \item Green dots represent singular trajectories exhibiting a first-order phase transition for $U_3$ along it. \end{enumerate}

Interestingly, and in contrast with our preliminary expectations in \cite{Lahoche6}, only these two potentials show a transition. Following the interpretation that is generally valid in this context \cite{QSG6}, one might be tempted to see the red phase as the analogue of a phase where ergodicity is broken dynamically (i.e., quantum dynamics fails to equilibrate throughout the space). Typical states in that phase belong to clusters whose number scales exponentially with $N$. This situation is reminiscent of what happens for the dynamical $p=2$ spin dynamics, where a weak ergodicity breaking inducing memory effects occurs without replica symmetry breaking. In the green phase, correlations appear between replicas, and the transition becomes discontinuous. One might be tempted to interpret this phase as associated with a "static transition," where the number of clusters becomes small, and a randomly selected pair of states will have a finite probability of belonging to the same cluster. Note that this phase is usually characterized by a breaking of the replica symmetry, which is particularly reflected in properties of ultrametricity. Here, we only have correlations between replicas that are assumed to reflect a specific breaking of ergodicity (which was suggested by the analytical results in \cite{Cugliandolo1,Cugliandolo2}), given that we do not take the formal limit $n\to 0$.

This work obviously needs to be further developed and suffers from many limitations in its applications, notably related to the fact that the potentials we work with are perturbative. A complementary analytical work, based on a completely 2PI approach, is currently being developed and should allow us to define a confidence zone in the approximations we have used here, whose simplicity would make it a potentially interesting tool for more complicated quantum problems than the academic model we have used here as a benchmark.

\appendix

\section{Flow equations}\label{AppA}

The flow equations for $u_2$ arbitrary large have been derived in \cite{Lahoche1}. We recall them here without detail:

\begin{equation}
\dot{{u}}_2= -\frac{{u}_4}{18} I_2(u_2)\,,
\end{equation}
\begin{equation}
\dot{{u}}_4=- \frac{{\tilde{u}}_6 }{15\pi}\tilde{I}_2(u_2)-\frac{{u}_6 }{30} I_2(u_2)+ \frac{{u}_4^2}{6}I_3(u_2)\,,
\end{equation}
\begin{equation}
\dot{{{u}}}_6= \frac{144{u}_4{u}_6}{5}I_3(u_2)+\frac{8{u}_4{\tilde{u}}_6}{5\pi}\tilde{I}_3(u_2) -\frac{5{u}_4^3}{9}I_4(u_2) \,,
\end{equation}

where:

\begin{equation}
\dot{R}_k(p^2):= \frac{d}{dt} R_k(p^2)\,,
\end{equation}
$t=\ln(k)$, and:

\begin{equation}
I_n(u_2):=\int dp^2\, \frac{\rho(p^2) \dot{R}_k(p^2)\sqrt{\pi}\,  \Gamma \left(n-\frac{1}{2}\right)}{\Gamma (n)(p^2+u_2+R_k(p^2))^{n-\frac{1}{2}}}\,,
\end{equation}

\begin{equation}
\tilde{I}_n(u_2):=\int dp^2\, \frac{\rho(p^2) \dot{R}_k(p^2)}{(p^2+u_2+R_k(p^2))^{n}}\,.
\end{equation}

\end{document}